\documentclass[journal=jctc ,manuscript=article]{achemso}
\usepackage{soul}      % pour \hl
\usepackage{xcolor}    % couleurs
\sethlcolor{yellow}    % jaune par défaut
\usepackage{upgreek}

\usepackage[version=3]{mhchem} % Formula subscripts using \ce{}

\author{El Hassane Lahrar}
\affiliation[1]{Sorbonne Universit\'e, CNRS, Physicochimie des \'Electrolytes et Nanosyst\`emes Interfaciaux, F-75005 Paris, France}
\alsoaffiliation[2]{Université de Toulouse, Toulouse INP, CNRS, CIRIMAT, Toulouse, France}
\alsoaffiliation[3]{R\'eseau sur le Stockage \'Electrochimique de l'\'Energie (RS2E), F\'ed\'eration de Recherche CNRS 3459, HUB de l'\'Energie, Rue Baudelocque,  80039 Amiens, France}
\email{el-hassane.lahrar@utoulouse.fr}

\author{Mathieu Salanne}
\affiliation[1]{Sorbonne Universit\'e, CNRS, Physicochimie des \'Electrolytes et Nanosyst\`emes Interfaciaux, F-75005 Paris, France}
\alsoaffiliation[3]{R\'eseau sur le Stockage \'Electrochimique de l'\'Energie (RS2E), F\'ed\'eration de Recherche CNRS 3459, HUB de l'\'Energie, Rue Baudelocque, 80039 Amiens, France}

\author{Rudolf Weeber}
\affiliation[4]{Institute for Computational Physics, University of Stuttgart, Allmandring 3, 70569 Stuttgart, Germany}

\author{C\'eline Merlet}
\affiliation[2]{Université de Toulouse, Toulouse INP, CNRS, CIRIMAT, Toulouse, France}
\alsoaffiliation[3]{R\'eseau sur le Stockage \'Electrochimique de l'\'Energie (RS2E), F\'ed\'eration de Recherche CNRS 3459, HUB de l'\'Energie, Rue Baudelocque,  80039 Amiens, France}
\email{celine.merlet@utoulouse.fr}

\title{LPC3D: An Enhanced Parallel Software for Large-Scale Simulation of Adsorption in Porous Carbons and Supercapacitors}


\begin{document}

\begin{tocentry}

\includegraphics[scale=0.13]{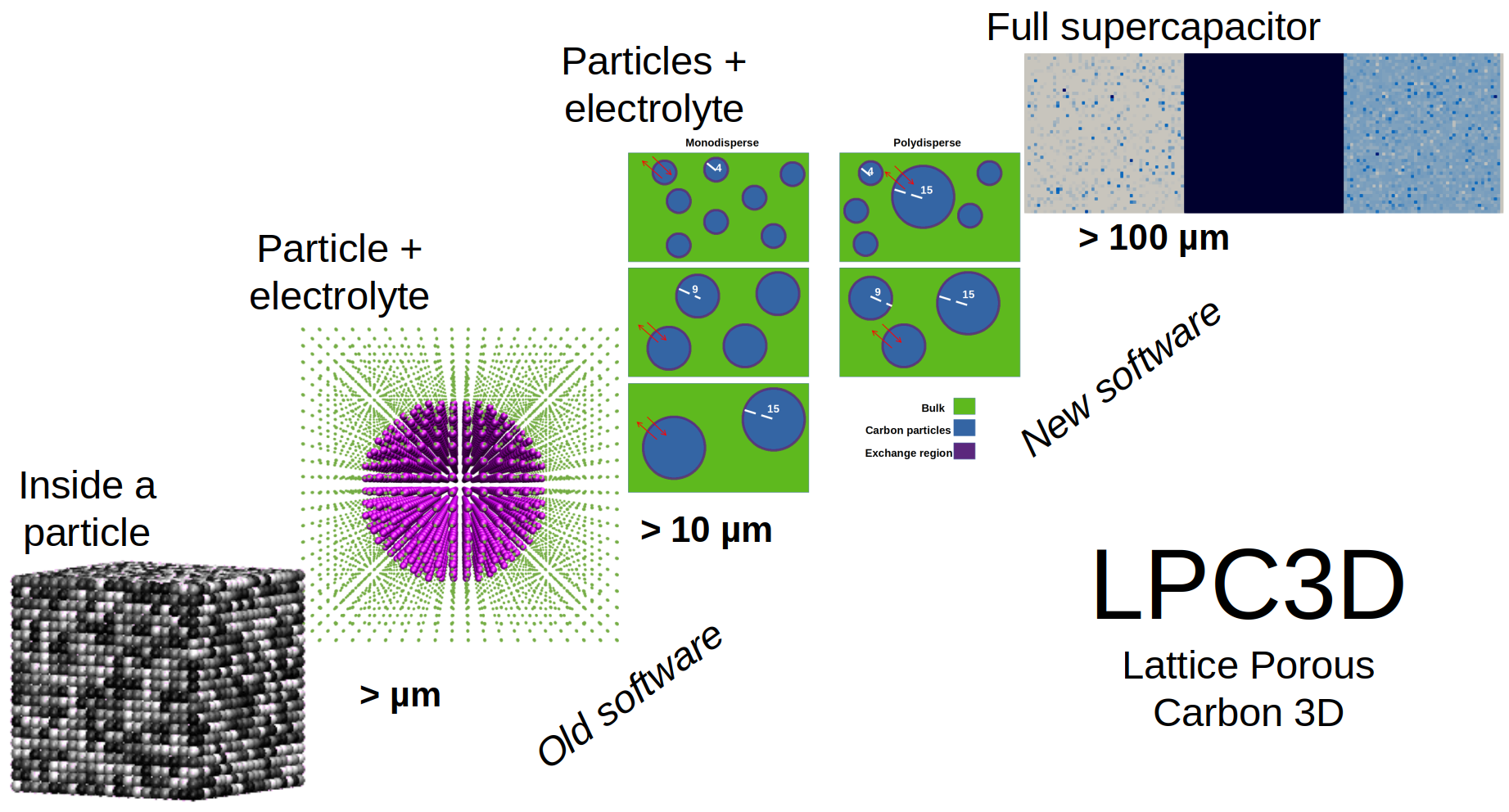}

\end{tocentry}

\begin{abstract}
Simulations of electrochemical double layer capacitors based on porous carbon electrodes, energy storage systems which accumulate and release energy through reversible ion adsorption at electrode/electrolyte interfaces, are often performed at the microscopic scale, using molecular dynamics. Such simulations provide crucial information to understand the adsorption of ions and the effect of confinement on some electrochemical properties. However, their computational cost limits the size of the systems studied to a few nanometers and a few pores while experimental materials are highly heterogeneous with a distribution of particle and pore sizes. LPC3D is a software designed for mesoscopic simulations of porous carbon particles and carbon-based supercapacitors which allow for the inclusion of such heterogeneity. The code calculates quantities of adsorbed ions, diffusion coefficients and NMR spectra of ions / molecules adsorbed in porous carbon matrices. In this work, we report on a new implementation of LPC3D, written in Python using the {\tt{PyStencils}} module which can generate optimized C++ and CUDA code. This implementation is parallel, can be run on CPU and GPU, and allows one to simulate systems going from a single carbon particle to a supercapacitor with hundreds of micrometers in length. Here, we apply the new implementation of LPC3D to the simulation of supercapacitors with porous carbon electrodes represented as monoliths or carbon films to investigate the influence of the microstructure on the resulting adsorption and spectroscopic properties. 
\end{abstract}

\section{Introduction}

Electrochemical double-layer capacitors (EDLCs), also known as supercapacitors, are energy storage systems characterized by their high power densities, long cycle lifetimes, and wide operating temperature range~\citep{Zhong15,Beguin14,Brandt13,Lewandowski10}. In these systems, charge storage is based on the voltage-driven electrosorption of ions from an electrolyte onto the electrode surfaces~\citep{Simon-Gogotsi,Forse16,Salanne16}. Owing to their exceptionally high specific surface area, good electronic conductivity, chemical stability, and ease of synthesis, porous carbon materials are the electrodes of choice \citep{Simon13}. Numerous studies have been conducted to better understand the parameters influencing the capacitance and, subsequently, the energy stored in these systems. Key features include pore size and its correlation with ion size, the curvature of the electrodes, the degree of ordering and the potential of zero charge~\citep{Chmiola06,Raymundo-Pinero06,Merlet12,Feng-curv,Liu24,Ge25}. 

Given the pivotal role of carbon structure in supercapacitor performance, understanding ion behavior under confinement within these porous structures is essential. In this regard, Nuclear Magnetic Resonance (NMR) emerges as a powerful technique for investigating the interactions and dynamics of ions within pores~\citep{Forse21,Asres25}. The non-invasive and nucleus specific nature of NMR makes it particularly valuable for studying confined species. As a quantitative method and thanks to the distinct chemical shifts exhibited by species in the bulk (`ex-pore') versus those adsorbed within pores (`in-pore'), NMR allows for the determination of the quantities of adsorbed ions and solvent molecules, at various potentials and regardless of their charge~\citep{john2014,john2016}. While NMR is a very powerful technique in this regard, dynamic exchange between environments can lead to peak merging and complex lineshapes in NMR spectra~\citep{Lyu24}, difficult to interpret without complementary theoretical approaches.

On the theoretical side, most simulations of supercapacitors have been performed at the microscopic scale,~\citep{Jeanmairet22} using molecular dynamics software such as LAMMPS~\citep{LAMMPS}, MetalWalls~\citep{marinlafleche2020a,coretti2022a} and ESPResSo~\citep{Espresso, Espresso4.0}. Such simulations aim at understanding the adsorption of ions and the effect of porosity on some electrochemical properties~\citep{Merlet12,Feng2011,Kondrat23,Khlyupin25,cummings2024}. One challenge associated with molecular dynamics simulations of supercapacitors is the generation of accurate atomistic models of complex, disordered carbons used as electrodes. A number of methods have been proposed in the past to obtain such disordered structures, including Reverse Monte Carlo~\cite{Thomson00}, Hybrid Reverse Monte Carlo~\citep{Farmahini13,Palmer09b}, Quench Molecular Dynamics~\citep{Palmer10,Thompson17,Deringer18}, and mimetic approaches related to experimental conditions~\citep{deTomas17,Schweiser17,VallejosBurgos23}. The atomistic carbon structures generated in such ways are of great value, as they permit the investigation of microscopic phenomena at the molecular level. However, this increased accuracy comes at the expense of a significantly higher computational cost and more difficult characterization compared to slit pore models~\citep{Bhatia16,Farmahini15}, which prevents the use of molecular simulations to screen porous carbons for supercapacitor applications. 

Another limitation of molecular simulations is the restricted range of time and length scales that can be probed, typically in the order of a few nanoseconds and a few nanometers, which is considerably smaller than the typical experimental values of several micrometers for the electrode width and  of milliseconds to seconds for the charging time. It is also well known from experiments that materials are highly inhomogeneous, exhibiting broad pore size and particle size distributions. While studies on the effect of particle size are rare, it was shown for example that sieving influences the NMR spectra measured~\citep{Cervini19} and that particle sizes  influence density, surface morphology, and specific capacitance~\citep{Garcia19,Taer18}. Molecular simulations, where electrode sizes are a few nanometers, only allow for the inclusion of a few pores which are insufficient to simulate electrodes and supercapacitors at larger scales. 

To be able to simulate systems large enough to include realistic pore size and particle size distributions, the Lattice Porous Carbon 3D~\citep{Merlet2015} software (LPC3D) has been developed. It is based on a lattice gas model and aims at simulating ion adsorption and diffusion in porous carbon based systems, such as supercapacitors. The code calculates quantities of adsorbed ions, diffusion coefficients, and NMR spectra relying on information obtained from molecular simulations as its main input. In particular, Nucleus Independent Chemical Shifts (necessary for the calculation of NMR spectra) are obtained from Density Functional Theory and adsorption energies in pores of different sizes are taken from classical molecular dynamics. The first version of LPC3D was a serial code written in~C. It allowed routine simulations of systems with around 100,000 sites, which correspond to a dimension of approximately 1~$\upmu$m$^{3}$, i.e. the size of a single - relatively small - particle, much smaller than an electrode or a supercapacitor device~\citep{belhboub2019, anagha2021, anagha2022, anagha2023, lahrar2024}.

In this work, based on the original LPC3D software, we have implemented and report on a new parallel version. With this new implementation, systems with several hundred million sites can be routinely simulated. As a consequence, the program is able to simulate a variety of systems, ranging from a single carbon particle ($\sim$~1~$\upmu$m$^{3}$) to a full supercapacitor ($\sim$~1000~$\upmu$m$^{3}$). The software has been released and can be found, along with a user manual and examples, at https://github.com/multixscale/LPC3D. It can also be installed directly using \texttt{pip install lpc3d}. Here, we apply it to the simulation of two supercapacitor systems with either monolithic or film-like electrodes.

\section{Methods}

\subsection{Description of the model}

\begin{figure}[ht]
\centering
  \includegraphics[scale=0.4]{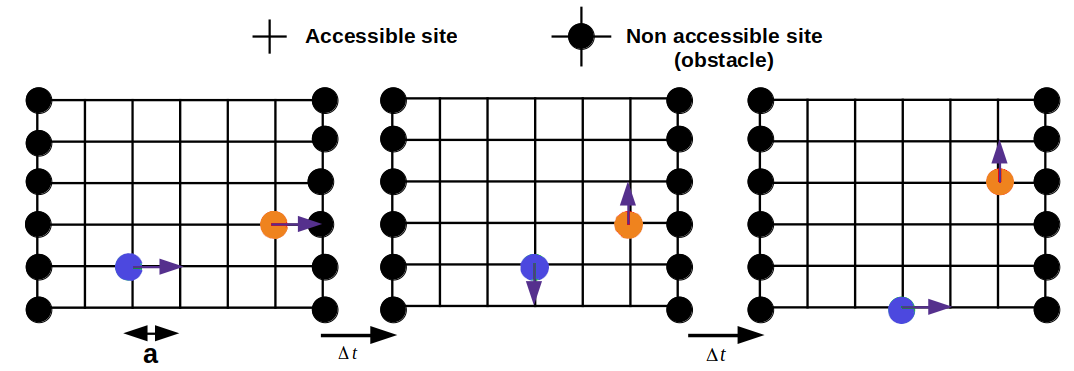}
  \caption{Illustration of a 2D lattice gas model. The lattice, characterized by a lattice parameter $a$, is divided between two types of sites: accessible (fluid) and not accessible (porous matrix). If at a given timestep, a particle velocity points towards an obstacle, then at the next timestep, that particle remains at its original position. Otherwise, it moves to the neighboring site.}
  \label{fgr:2D_mod}
\end{figure}

The model implemented here has already been described in detail in a previous article~\citep{Merlet2015} but we recall here its main aspects for a better readability. The simplest lattice-gas models employed to simulate diffusion phenomena represent the mobile species as non-interacting particles executing kinetic Monte Carlo displacements on a lattice comprising both accessible and inaccessible sites. A two-dimensional representation of this type of model is given in Figure~\ref{fgr:2D_mod}. In the present case, particles can be ions or solvent molecules.
The lattice model is characterized by two key quantities: the lattice parameter $a$, which represents the distance between two lattice nodes, and the timestep $\Delta t$, which corresponds to the typical time it takes the probe particle to diffuse over a distance $a$. The spatial distribution of excluded lattice sites provides the geometry of the solid matrix.

In the simplest form of the lattice-gas model, at each timestep, particles have an equal probability of attempting to move to any of their nearest neighboring sites (six in total for a simple lattice in three dimensions). If the chosen neighboring site is accessible, the particle moves to this new position; otherwise, it remains at its original location. For the systems we are interested in, to simulate the diffusion of solvent molecules and electrolyte ions within a porous carbon matrix, it is essential to consider that accessible sites exhibit varying energy levels (more precisely, free energies), resulting in unequal site occupancy. These energies reflect the fact that not all pore sizes are as favorable for adsorption. An energy value $E_{i}$ must then be assigned to each site $i$. Furthermore, when simulating NMR lineshapes, it is necessary to account for variations in the local magnetic field (due to different carbon environments), which leads to site-specific resonant frequencies $\nu_{i}$.

The diffusion of species between lattice sites is determined by a Metropolis acceptance rule through which a transition from site $i$ to site $j$ follows the probability: 
\begin{equation}\label{probability}
P(i\to j) = \left\{
\begin{array}{cc}
\exp\left ( \frac{-(E_j-E_i)}{{\rm k_B}T} \right ) & \mathrm{if\ } E_j>E_i \\
1 & \mathrm{if\ } E_j\leq {E_i}
\end{array}
\right.
\end{equation}
where $E_i$ is the energy assigned to site $i$, ${\rm k_B}$ is the Boltzmann constant and $T$ is the temperature of the system. A transition from a site $i$, characterized by a higher energy level, to a site $j$, characterized by a lower energy level, will always take place. The probability of the reverse transition decreases as the difference between the energies of the two sites, $E_i - E_j$, increases. With these rules, the probability of a particle to visit site $i$ is given by the Boltzmann distribution: 
\begin{equation}
\rho_{i}= <\rho> \times \frac {\exp\left( \frac{-E_i}{{\rm k_B}T} \right )} {\sum_{j}\exp\left( \frac{-E_j}{{\rm k_B}T}\right )}
\label{eq:Boltzmann-densities}
\end{equation}
where $\rho_{i}$ is the average density of particles at site $i$ and $<\rho>$ is the density averaged over all lattice sites. It is worth noting that here we are not following individual particles moving across the lattice but average densities of particles on lattice sites.

\subsection{Diffusion coefficients}

Diffusion coefficients are usually determined with a Green-Kubo relation between the velocity autocorrelation function and the self-diffusion coefficient:
\begin{equation}
D = \int_{0}^\infty \frac{1}{d} <v(0) \cdot v(t)>\cdot dt
\label{diff}
\end{equation}
where $D$ is the diffusion coefficient, $d$ is the dimensionality of the system, $v(0)$ is the velocity at time $t = 0$, $v(t)$ is the velocity at time $t$, and $\langle ...\rangle$ denotes an average over all particles and all trajectories. Using the Kohn approach to discretize equation~\ref{diff}, we obtain the following relation~\citep{Merlet2015}:
\begin{equation}
D = \frac{1}{2\Delta t}<(\Delta x)^2> + \frac{1}{\Delta t} \sum_{j=2}^\infty <\Delta x_{1} \Delta x_{j}>
\label{diff_discrete}
\end{equation}
where $\Delta x_{i}$ indicates the displacement of a particle in the $x$ direction at timestep $i$, and $\langle ...\rangle$ indicates an average over all sites taken into consideration. 

One way to estimate diffusion coefficients is to generate random trajectories for several particles starting at various points in the lattice. This method is extremely inefficient as it requires the construction of enough trajectories for a heterogeneous system to produce accurate statistics. Here, however, we employ an alternative strategy: the so-called moment-propagation method, which was proposed by Frenkel and effectively used in several investigations of particle dynamics in restricted systems~\citep{Levesque13,Frenkel87,Rotenb08}.

The moment-propagation method is a recursive computational scheme that enables the sampling of all possible trajectories of diffusing particles, rather than only a subset. The computational effort required by this method scales linearly with the product of the simulation time $t$ and the number of lattice sites $M$, expressed as $t$ $\times$ $M$. In contrast, employing a non-recursive scheme to sample all trajectories would result in computational effort that scales exponentially with time, specifically as $z^t M$, where $z$ denotes the coordination number of the lattice. 

In equation~\ref{diff_discrete}, the first term in the expression for diffusion coefficients is equal to $\frac{1}{2\Delta t}$ times the mean-square displacement of a particle in the $x$-direction during one timestep, it can be simply calculated as:
\begin{equation}
\frac{1}{2\Delta t} <(\Delta x)^2> = \frac{1}{2z\Delta t} \sum_{j=1}^{z} p_{\text{acc}}(i \rightarrow j) a_x^2(j).
\end{equation}
In this context, $a_x$ denotes the $x$-component of the vector connecting a lattice site $i$ to its $j$-th neighboring site. The summation extends over all sites $j$ adjacent to $i$, noting that $p_{\text{acc}}(i \rightarrow j) = 0$ if site $j$ is occupied by an obstacle. 

The diffusion coefficient $D$ in a homogeneous fluid with uncorrelated successive particle jumps is determined by:
\begin{equation}
D = \frac{1}{2\Delta t} <(\Delta x)^2>
\end{equation}
In case the probability of jumping to a neighboring site is one, the equation for $D$ is equal to $\frac{a^2}{2d\Delta t}$, where $a$ represents the lattice spacing and $d$ the dimension of the system. We can nevertheless decrease the probability that a particle will make a jump using an additional parameter $\alpha$ (comprised between 0 and 1). If we write $1 - \alpha$ as the chance that a particle remains at the same site, then: 
\begin{equation}
D = \alpha \frac{a^2}{2d\Delta t}
\end{equation}
In systems where diffusivity varies with position, a factor $\alpha_{ij}$ can be defined for each link between neighboring lattice sites $i$ and $j$. It is important to ensure that if the probability of jumping from site $i$ to site $j$ is reduced, the probability of jumping from $j$ to $i$ must be reduced by the same factor, \emph{i.e.} we follow detailed balance. If not, the equilibrium distribution over lattice sites would be different from the Boltzmann distribution according to site energies.

\subsection{NMR spectra}

NMR spectroscopy measures the nuclear magnetic response of a sample when a radio frequency pulse perturbs its equilibrium. Following this pulse, the sample transverse magnetization decays. The NMR signal observed during this decay is known as the Free Induction Decay (FID). The NMR spectrum is generated by Fourier transforming this FID signal. In LPC3D, a similar process is used to determine the NMR signal. In the systems of interest here, nuclei are in a heterogeneous sample and encounter various local magnetic fields, resulting in varying Larmor frequencies. As the nuclei diffuse, their environment and resonance frequencies change. The FID signal is a superposition of signals from all excited nuclei in the sample and can be calculated as follows:
\begin{equation}  
G(t)=\langle {\rm e}^{{\rm i}\int_0^t 2\pi\nu_i^0(t').dt'}\rangle
\label{NMR-signal-equation}
\end{equation}
where $\nu_i^0(t)$ is the Larmor frequency corresponding to spin $i$ at time $t$, and $\langle ...\rangle$ denotes an average over all spins. The spectrum is then obtained by Fourier transforming this signal
following:
\begin{equation}  
F(k) = \int_{-\infty}^{\infty} dt G(t) e^{-2i\pi k t} \, dt
\end{equation}
To use the moment propagation method, the same strategy as calculating diffusion coefficients has been carried out to discretize and estimate the expression. A full description of the implementation using the moment propagation approach can be found in the work by Merlet~\emph{et~al.}~\citep{Merlet2015}.

\section{Software implementation and performance}

\subsection{Implementation using PyStencils}

\begin{figure}[ht]
\centering
  \includegraphics[scale=0.28]{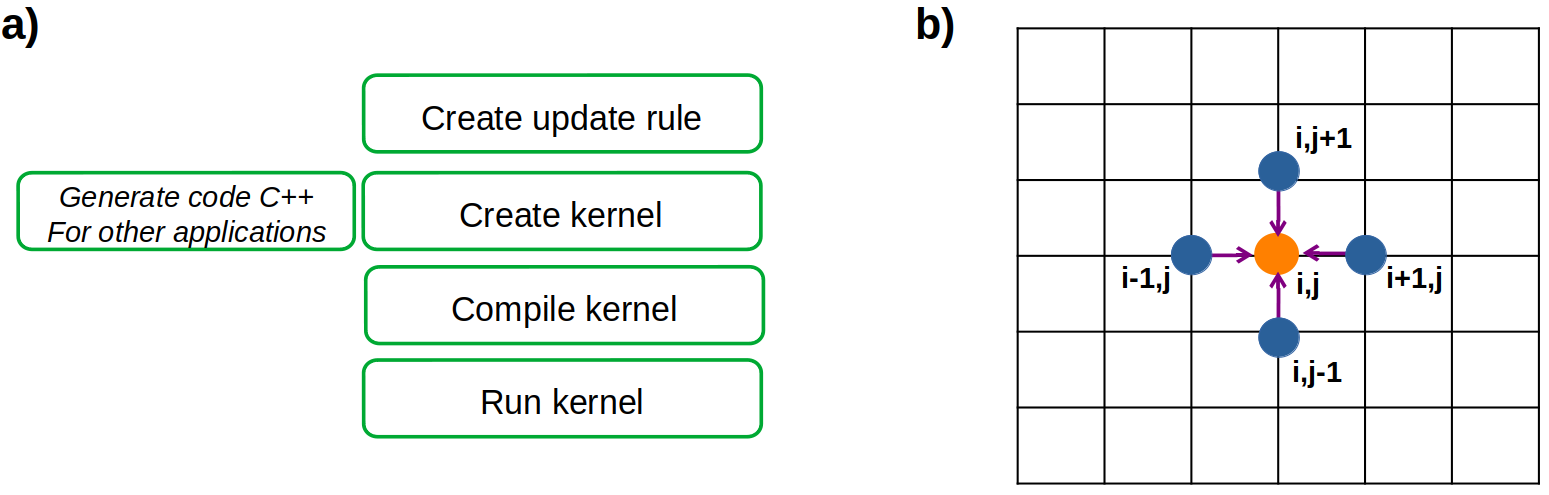}
  \caption{a) Different processes used in {\tt{PyStencils}} to create and execute a kernel. b) Illustration of a lattice gas model update rule in 2D: The central site (in red) receives contributions from its four neighboring cells (in green), showing how particles and functions propagate to adjacent cells based on the update rules defined in the model.}
  \label{fgr:pystencilss}
\end{figure}

{\tt{PyStencils}}~\citep{Bauer19,Ernst23} is a Python-based framework stencil methods. A stencil method, such as the lattice gas model discussed in this article, is a numerical scheme that updates values at a lattice site using a fixed pattern of neighboring lattice sites. {\tt{PyStencils}}, based on update rules provided in terms of mathematical symbols, automatically generates, optimizes and runs the code to apply these mathematical rules on the lattice. This can be done on the CPU, including OpenMP and MPI parallelism as well as on the GPU using largely the same Python code, providing flexibility and performance portability. The process is illustrated in Figure~\ref{fgr:pystencilss}a The process starts with defining the update rule for a particular stencil operation (defined symbolically). Then, followed by creating a corresponding kernel (computational routine) that encapsulates this rule. This kernel is then compiled and optimized into C++ or CUDA code (depending on the target hardware), which can be run directly from Python using NumPy arrays to store the lattice data.

We illustrate the principle of a stencil calculation on the NMR signal calculation on a two-dimensional lattice. The equation for the NMR signal,
\begin{equation}  
G(t)=\langle {\rm e}^{{\rm i}\int_0^t 2\pi\nu(t').dt'}\rangle
\end{equation}
can be discretised into:
\begin{equation}
G(n\Delta t)=\langle {\rm e}^{{\rm i} \Delta t\sum_{n=1}^N 2\pi\nu(n)}\rangle
\end{equation}
This function can be calculated from functions defined in all sites $(i,j)$, $g_{i,j}(n\Delta t)$, such that:
\begin{equation}
G(n\Delta t) = \sum_{i,j} g_{i,j}(n\Delta t)
\end{equation}
where the sum runs over all sites of the lattice. 
For a given site $(i,j)$, the function $g_{i,j}$ is computed recursively with:
\begin{eqnarray}
g_{i,j}(n\Delta t) = e^{{\rm i} \Delta t \nu_{i,j}} [ & P((i+1,j)\to (i,j))\times g_{i+1,j}((n-1)\Delta t)) +  \nonumber\\ 
 & P((i-1,j)\to (i,j))\times g_{i-1,j}((n-1)\Delta t)) + \nonumber\\
 & P((i,j+1)\to (i,j))\times g_{i,j+1}((n-1)\Delta t)) +  \\
 & P((i,j-1)\to (i,j))\times g_{i,j-1}((n-1)\Delta t))  ]. \nonumber
\end{eqnarray}
This rule captures the evolution of the local NMR signal when particles are moving to neighboring cells. Figure~\ref{fgr:pystencilss}b illustrates this process where the central site (i,j), in red, receives contributions from its four neighbors, in green, based on the update rule. Such stencil calculations can be setup for many functions in the current model.

{\tt{PyStencils}} is versatile and can be adapted to a wide range of scientific simulations that require stencil computations, such as fluid dynamics, heat transfer, and electromagnetics. The ability of the framework to symbolically define operations and optimize them for different hardware architectures makes it an invaluable tool to bridge the gap between high-level model development and low-level performance optimization.

\subsection{General work-flow of the software}

The workflow adopted follows a structured sequence of steps. This systematic approach ensures that the simulations are not only computationally efficient but also adaptable to different hardware configurations, making the software versatile for execution on both standard desktop environments and High-Performance Computing (HPC) clusters.

\begin{figure}[ht!]
\centering
  \includegraphics[scale=0.28]{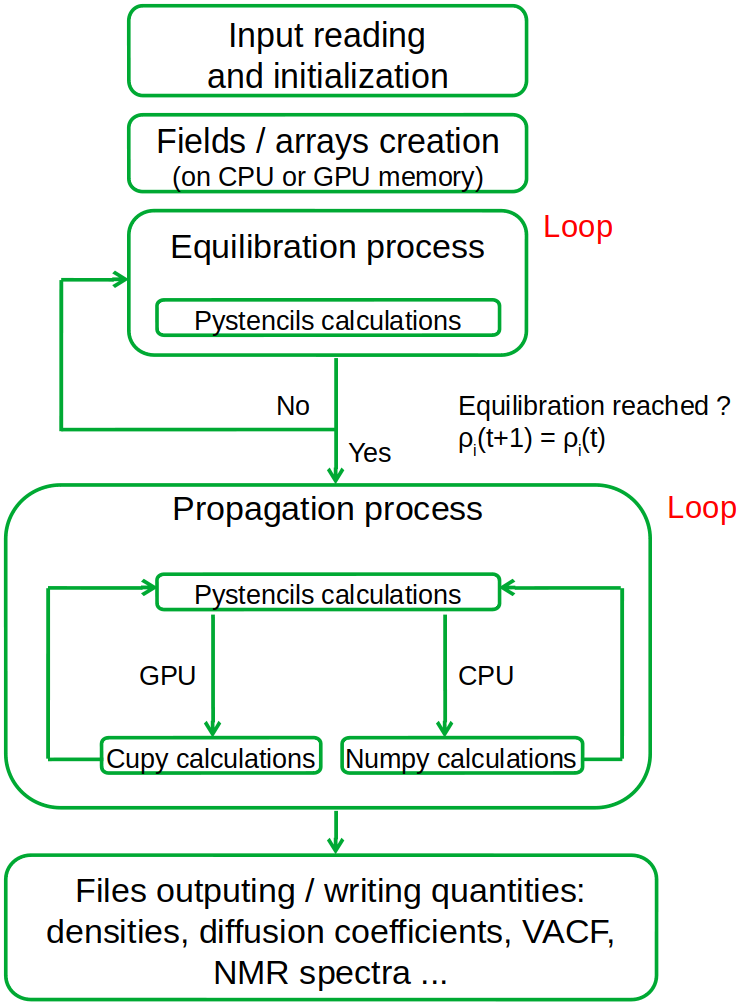}
  \caption{Steps involved in conducting a calculation with the lattice-gas model}
  \label{fgr:wrk}
\end{figure}

Figure~\ref{fgr:wrk} illustrates the main steps involved in conducting a calculation with the lattice-gas model. The first steps are related to the reading of input files which will be used to parametrise the model, according to the representation of a chosen systen (given pore size distribution, electrolyte species, NMR parameters). The fields and arrays necessary for the calculation are then created. Following this step, a period of equilibration can be realised or not depending on the user settings. In particular, the densities of electrolyte species on sites can be initially set equal to Boltzmann averages (see equation~\ref{eq:Boltzmann-densities}) or to any user-defined values. The equilibration stops if either the densities become constant or if the maximum number of iterations to reach equilibrium is attained. Next, the propagation step of the moment-propagation approach is conducted for all required quantities. Finally, properties of interest are written.

These various steps are detailed below.

\begin{enumerate}
    \item \textbf{Input reading and initialization}\\
    The software begins by reading the input file, which defines all the simulation parameters, such as system dimensions (i.e. number of lattice sites in the three dimensions), timestep, duration of the simulation, boundary conditions, and the user’s choice of computational environment. The software then automatically configures tables and fields compatible with a CPU$\_$OpenMP parallel execution or a GPU-based approach using CUDA. During initialization, ghost layers are created to handle periodic boundary conditions. For non-periodic boundaries, Dirichlet-type conditions are imposed by fixing field values in ghost layers, ensuring constant values at the domain boundaries. Finally, different blocks are defined as either electrode or bulk regions, providing the necessary structural framework for simulating complex systems like supercapacitors.

    \item \textbf{Fields / arrays creation}\\
    Based on the initialization settings, the software uses {\tt{PyStencils}} to generate fields and arrays corresponding to various physical quantities, such as density, energies and velocities. These fields represent the spatial discretization of the simulation domain and are crucial for lattice-based calculations. {\tt{PyStencils}} ensures efficient data handling and parallel updates, allowing the simulation to adapt to different configurations based on user input.

    \item \textbf{Equilibration process}\\
    The equilibration phase employs {\tt{PyStencils}} to update local densities at each lattice site until a steady-state distribution is achieved (or the maximum number of iteration is reached). In the context of supercapacitors, this step can for example be used to simulate the fluid entering or leaving the nanopores of the electrode when a potential is applied. To check if equilibrium is reached, the densities at time $t$ ($\rho_{i}(t)$) are compared to those at the next time step ($\rho_{i}(t+1)$). If the difference is below a defined threshold, the system is considered equilibrated and the software moves on to the propagation phase. Otherwise, the loop continues until convergence is reached (or a maximum number of steps is attained).

    \item \textbf{Propagation process}\\
    Once the optional equilibration is completed, the software proceeds to the propagation phase. Dynamic quantities are calculated based on the equilibrated lattice. This stage involves time-dependent simulations, where the system evolves using the chosen backend — either ``NumPy'' for CPU computations or ``CuPy'' for GPU-accelerated operations. During this phase, various properties such as diffusion coefficients, VACF, and FID signals (for subsequent generation of NMR spectra) are computed.

    \item \textbf{Files outputting / writing quantities}\\
    Once the propagation phase is over, the software generates comprehensive output files containing the calculated properties, including densities, diffusion coefficients, and VACF.  Additionally, NMR spectra are obtained through a Fourier transformation of the FID signals.

\end{enumerate}

\subsection{Scalability and performance metrics}

To evaluate its parallel efficiency, the LPC3D code was tested on CPU and GPU nodes on two HPC centers: CALMIP~\cite{CALMIP} and VEGA~\cite{VEGA}. The standard CPU nodes of CALMIP (Intel® Skylake 2.3 Ghz) have 36 cores and 192~GB of RAM. The GPU nodes have 36 cores, 4 GPUs (GPU Nvidia Volta V100) and 384~GB of RAM. The standard CPU nodes of VEGA (AMD EPYC 7H12) have 128 cores and 256~GB of RAM. The GPU nodes have 4 GPUs (NVIDIA A100) and  512~GB of RAM. 

System sizes going from 50$\times$50$\times$50 (125,000 sites) to 700$\times$700$\times$700 (343 million sites) have been considered. It is worth noting that with the initial serial implementation of LPC3D, only systems of size 30$\times$30$\times$30 (27,000 sites) were routinely simulated. Systems of sizes up to 50$\times$50$\times$50 could be simulated but not to explore a large amount of parameters. 

\begin{figure}[ht!]
    \centering    
    \includegraphics[scale=0.29]{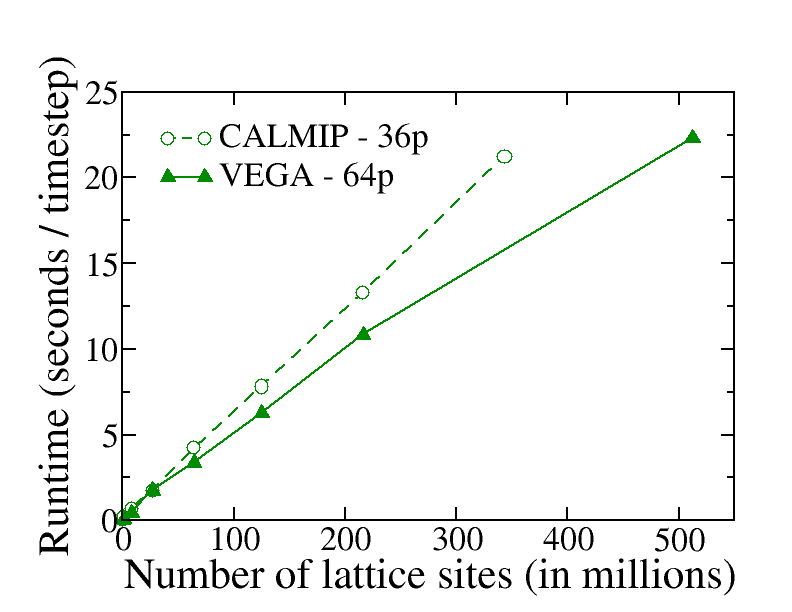}
    \includegraphics[scale=0.29]{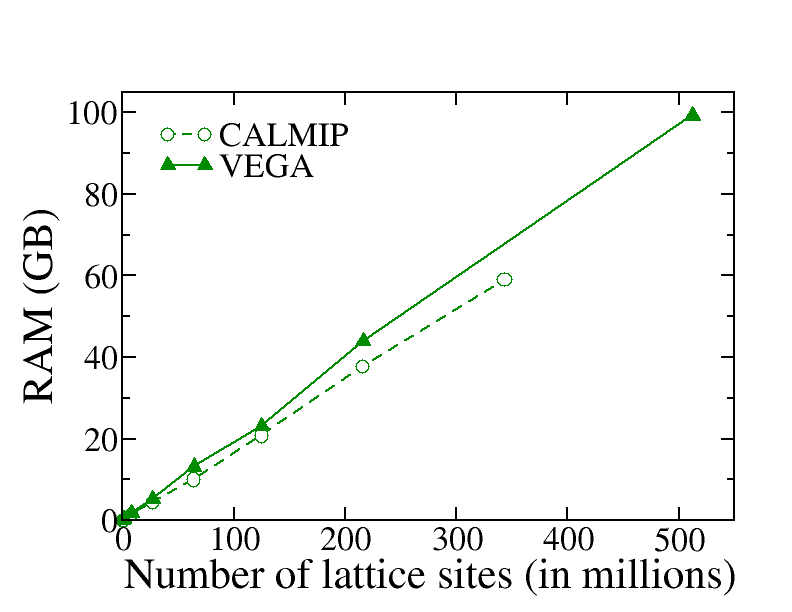}
    \caption{Runtime (in seconds per timestep of propagation) and memory usage for calculations done using the LP3CD code with various numbers of lattice sites, and conducted on CPU nodes exclusively}
    \label{fig:Figure4}
\end{figure}

Figure~\ref{fig:Figure4} shows the runtime (in seconds per timestep of propagation) and RAM usage for calculations done exclusively on CPU. The calculations reported here were done with 36~processes on CALMIP and 64~processes on VEGA. The runtime and memory usage confirm the linear scaling expected for the stencil operations conducted and for the information stored for all lattice sites. While the runtime is long for large numbers of lattice sites, several thousands of time steps can usually be simulated in a day. 

\begin{figure}[ht!]
    \centering    
    \includegraphics[scale=0.32]{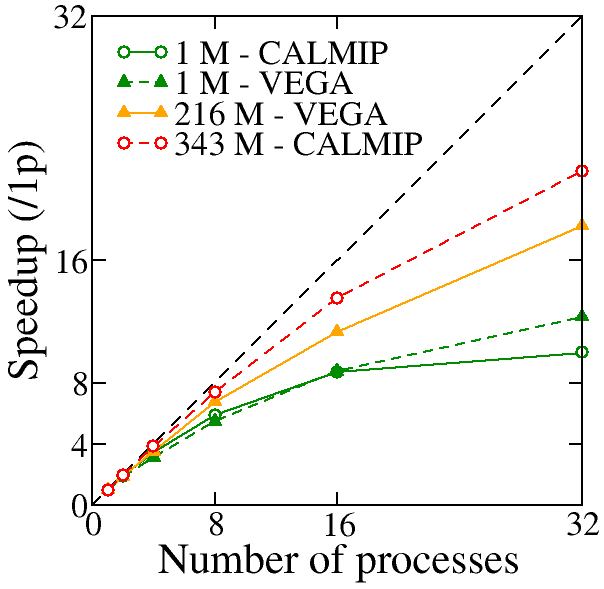}
    \hskip 20pt
    \includegraphics[scale=0.32]{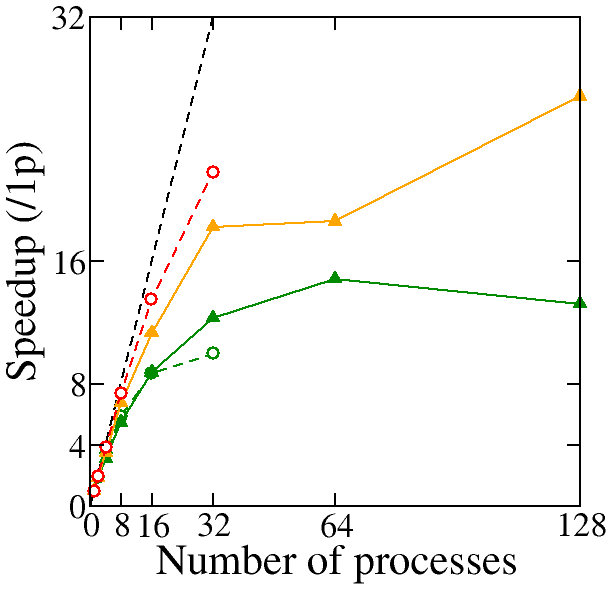}
    \caption{Evaluation of strong scaling performance of the LP3CD code. The dashed line indicate the ideal behaviour which would be observed if the code was perfectly parallel.}
    \label{fig:Figure5}
\end{figure}

Figure~\ref{fig:Figure5} shows the results of strong scaling tests. We report the speedup (i.e. acceleration) observed when the calculation is done on $N_p$ processes compared to 1 process. Calculations were conducted for three different system sizes (1, 216 and 343 million sites).
The speedup is good for numbers of processes up to 16 but is less ideal for larger numbers. The larger the system (i.e. the number of lattice sites), the better the scaling which is suggesting that the parallelization is not limited by the program itself but by the limited number of operations to perform.
A larger number of sites also improves the ratio between the time spent on calculation (which is parallel) and communication or remaining serial parts in the code usage. One limit to conduct simulations on larger systems is the RAM usage. It is worth noting that the system sizes considered here are sufficient to study the systems of interest and the performance of the code is thus satisfactory with no need for further improvement at this stage. 

\begin{figure}[ht!]
    \centering    
    \includegraphics[scale=0.29]{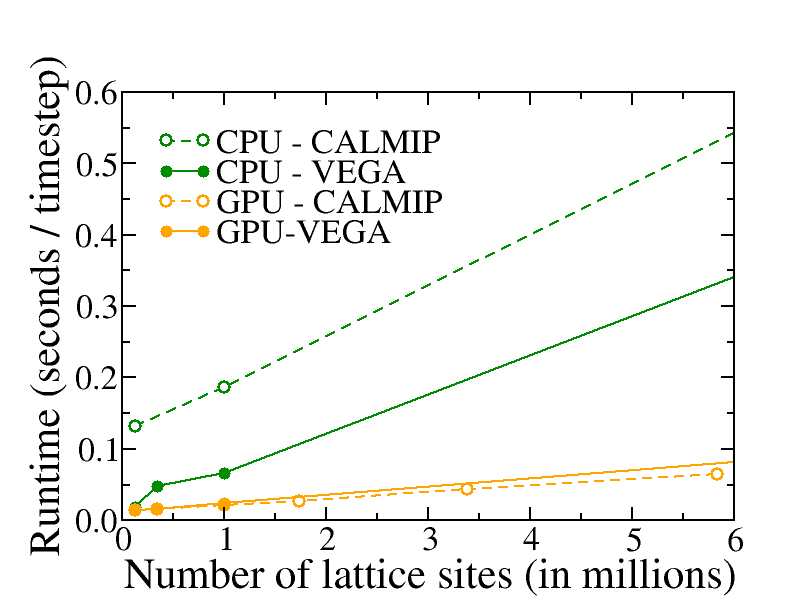}
    \includegraphics[scale=0.29]{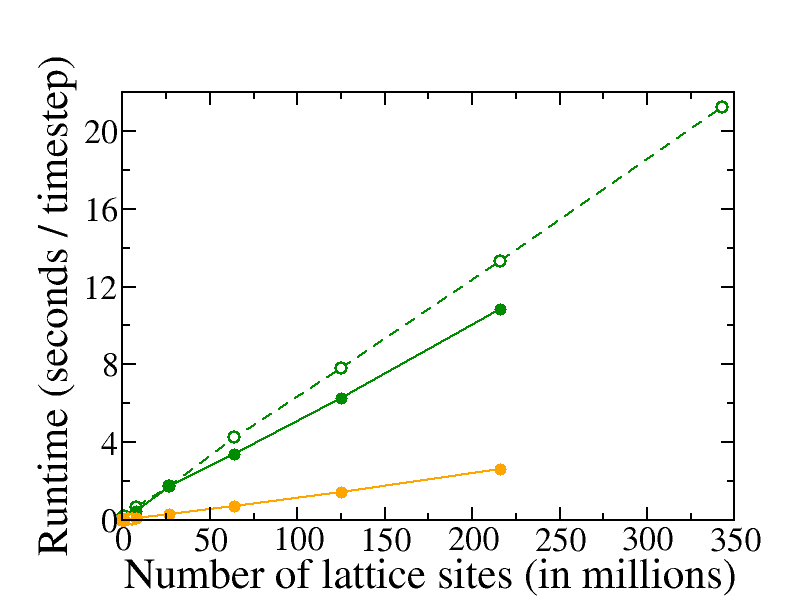}
    \caption{Runtime (in seconds per timestep of propagation) for CPU and GPU calculations using LPC3D for systems of various sizes.}
    \label{fig:Figure6}
\end{figure}

Figure~\ref{fig:Figure6} shows the comparison between CPU and GPU runtimes for systems of various sizes (Figure~\ref{fig:Figure6}a is a zoom of Figure~\ref{fig:Figure6}b on the lowest numbers of sites). For small numbers of lattice sites, the calculation times are similar. For larger system sizes, the calculation is much faster on GPU. The runtime for GPU also seems to be scaling linearly with the number of lattice sites as expected.

In summary, the performance of the implemented LPC3D software is very satisfactory. The aim was to go from systems of size 30$\times$30$\times$30, that corresponds to a dimension of approximately 1~$\mu\text{m}^3$, i.e. the size of a single - relatively small - particle, to systems where at least one dimension is a few hundreds of $\mu\text{m}$. With the current implementation, asymmetrical systems with around 1,000~$\mu\text{m}$ to 5,000~$\mu\text{m}$ in length and reasonable dimensions (sufficient considering periodic boundary conditions) can be simulated. Since the experimental thickness of electrodes is usually between 200 to 500~$\mu\text{m}$, the current implementation allows to simulate a full supercapacitor with two electrodes and bulk areas with realistic dimensions.

\section{Comparison of monolithic and film-like electrodes}

\begin{figure}[ht!]
    \centering    
    \includegraphics[scale=0.28]{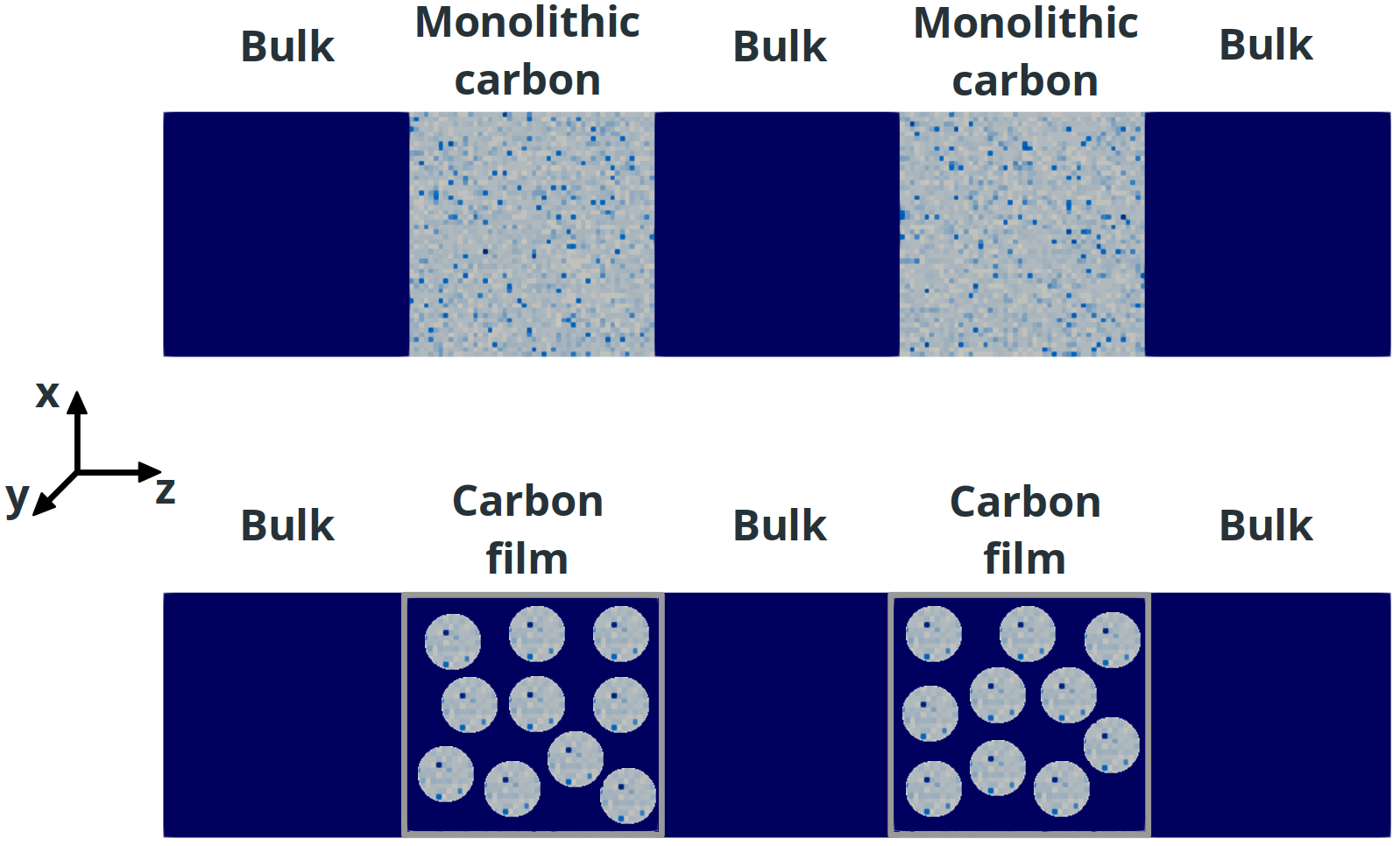}
    \caption{Illustration of the supercapacitors simulated: Two carbon electrodes are in contact with bulk electrolyte. Top: The electrodes are represented as monoliths in which all sites are pores of different sizes. Bottom: The electrodes are represented as films where carbon particles (here monodisperse) are separated by bulk-like regions. This second representation corresponds to the most common way of making electrodes experimentally. Different shades of blue in the carbon monoliths and particles correspond to different concentrations of anions (here at 0~V). The top figure is a result of simulation, visualised using paraview~\citep{paraview}. The bottom figure is a schematic illustration drawn from the top one.}
    \label{fig:Figure7}
\end{figure}

To illustrate the capabilities of the newly implemented software, we apply it to the simulation of full supercacitors with two types of electrodes: i)~monolithic electrodes where there are no particles but rather a single consistent porous matrix and ii)~film-like electrodes where porous carbon particles are separated by bulk-like regions. Figure~\ref{fig:Figure7} illustrates the two types of supercapacitors considered. In reality, experiments with both types of electrodes exist, but films are much more common~\citep{Huang16,Chmiola10,Moreno-Fernandez17}. The existence of unconfined diffusion between carbon particles in the film-like electrodes is expected to noticeably modify the dynamic properties of ions in the system and consequently the effective diffusion coefficients and NMR spectra.

To investigate these systems, we have conducted calculations for a 100$\times$100$\times$300 lattice, under two potentials, 0V and 2V. BMIM-BF$_4$ (1-butyl-3-methylimidazolium tetrafluroborate, 1.5~M) in acetonitrile was chosen as the electrolyte (i.e. the site energies $E_i$ and frequencies $\nu_i^0$ were parametrized according to DFT calculations and molecular simulations of this electrolyte in contact with carbon electrodes, see Sasikumar~\emph{et al.}~\citep{anagha2021} for details on the parametrization). In the case of the film, the particles are monodisperse with a radius equal to 9 lattice units and occupy 38\% of the total electrode volume. The pore size distribution is the same for monoliths and carbon particles, corresponding to the one of YP50F determined experimentally.~\citep{Cervini19} 

All simulations have been performed for 50,000 steps with a timestep of 5 $\mu$s. The residence time of ions in a given site, $\tau$, is defined to be 1~ms or 0.2~ms for all sites of the lattice, i.e. no distinction is made between sites in the bulk or in a particle. There are no periodic boundary conditions in the $z$ direction. The NMR spectra calculations are done considering the Larmor frequency of $^{19}$F with a 300~MHz spectrometer, i.e. 282~MHz. Indeed, $^{19}$F NMR is often used to probe electrolyte ions such as BF$_4^-$ or PF$_6^-$. These settings are adequate to see a complete decay of the free induction decay signal~\cite{Merlet2015,anagha2021,belhboub2019,anagha2023, lahrar2024}.

\subsection{Influence of electrode structure on quantities of adsorbed ions and diffusion coefficients}

\begin{figure}[ht!]
    \centering    
    \includegraphics[scale=0.24]{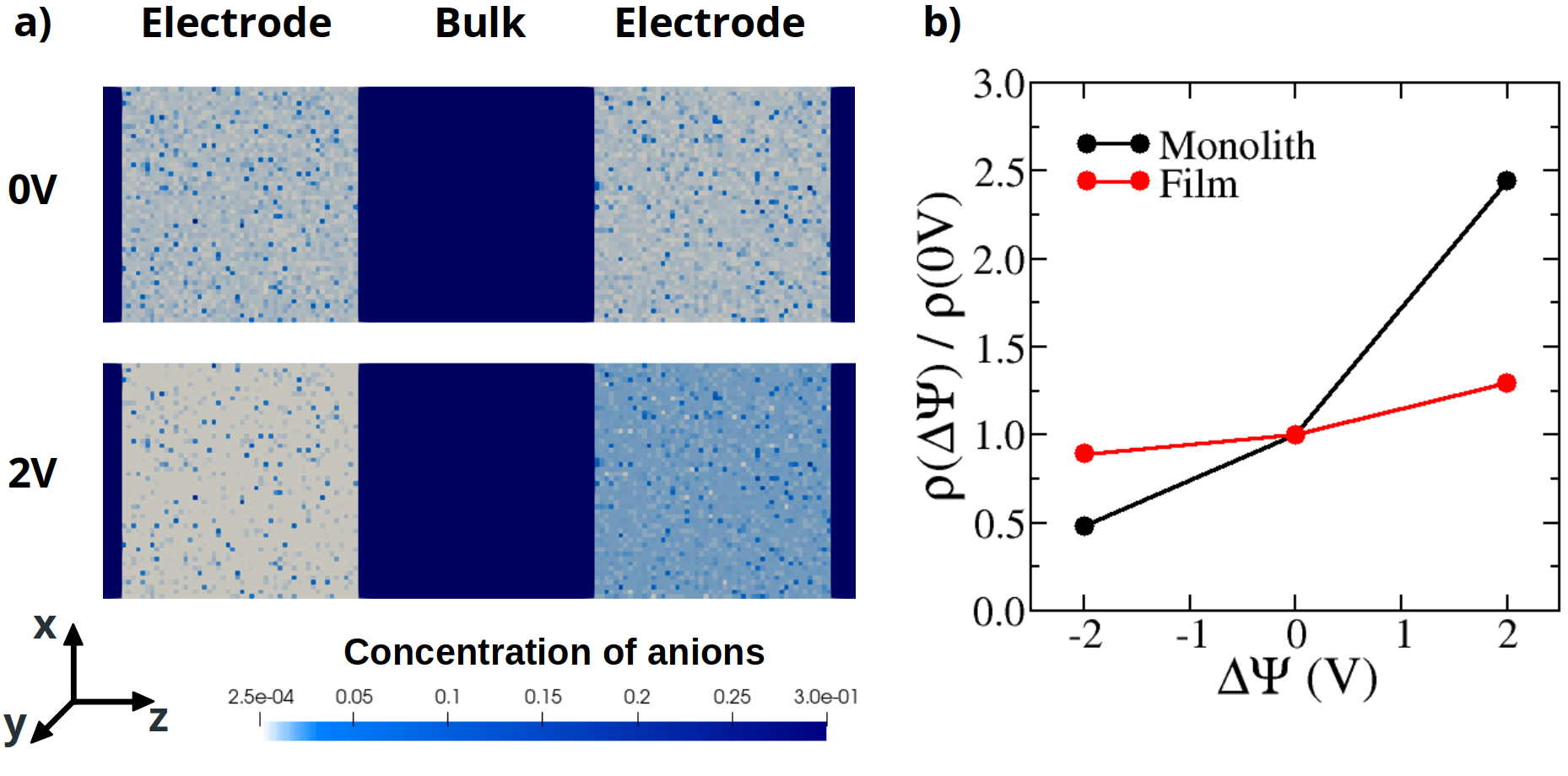}
    \caption{a) Illustration of the concentration of anions (in arbitrary units) in both electrodes of the monolith based supercapacitor at 0~V and 2~V. The pictures were made using paraview.~\citep{paraview} In the bulk, the concentration is maximum. In the pores of the electrodes, there are less ions, due to confinement. b) Normalized total quantity of anions, $\rho_{\mathrm{tot}}(\Delta\psi)/\rho_{\mathrm{tot}}(0\,\mathrm{V})$, versus applied potential for the two electrode morphologies.}
    \label{fig:density_tot}
\end{figure}

It is possible to simulate systems with an applied potential difference between the electrodes, by using energies obtained from molecular dynamics simulations conducted at different potentials, in which anions and cations enter or leave the electrodes. Here, we focus on the BF$_4^-$ anions. Figure~\ref{fig:density_tot}a shows the concentration of anions (in arbitrary units) in both electrodes for potentials of 0~V and 2~V. For both cases, the concentration is not homogeneous in the electrodes, as it depends on the pore size corresponding to each lattice site. It is very clear from the figure that anions tend to accumulate in the positive electrode and leave the negative electrode, consistently with the sign of the surface charge. 

To compare the two types of electrodes, the overall change in density between applied potentials of 0~V and 2~V is plotted in Figure~\ref{fig:density_tot}b. The monolith based supercapacitor exhibits a larger change in the quantity of adsorbed ions. It is expected as part of the film is not occupied by particles meaning that only part of the electrode will respond to the application of a potential. In practice, this can influence the capacitance of the supercapacitor as the bulk-like parts of the electrodes do not contribute to charge (and energy) storage. In practice, this can be related to experimental mass loading effects where sufficient powder have to be used in order to obtain films with good capacitive properties.

The LPC3D software also gives access to diffusion coefficients for different parts of the system, allowing for a comparison between electrode types and potentials applied. The results are given in Table~\ref{diff_coeff}. 
\begin{table}[h]
\begin{center}
\begin{tabular}{|c|c|c|c|}
\hline
 & 2~V / elec - & 0~V & 2~V / elec +  \\
\hline
Monolith & 0.89$\times D_{bulk}$ & 0.85$\times D_{bulk}$ & 0.93$\times D_{bulk}$\\
\hline
Film & 0.89$\times D_{bulk}$ & 0.93$\times D_{bulk}$ & 0.93$\times D_{bulk}$\\
\hline
\end{tabular}
\caption{Differential capacitances obtained for [BMI][PF$_6$], [BMI][BF$_4$] and [EMI][BF$_4$].}
\label{diff_coeff}
\end{center}
\end{table}
In all cases, the diffusion coefficients are very close to the ones of the bulk electrolyte. This is not surprising in the current setup, as the residence time is the same for all sites of the system, not differentiating between bulk or particle sites. In future works, it would be interesting to differentiate between those sites but here we focus on the differences observed due to electrode morphology. The diffusion is slightly faster in the film electrode at 0~V compared to the monolith but the diffusion coefficients are the same for the charged electrodes. This suggest that the diffusion is mainly modified by the ion concentration in this case. Heterogeneity in pore filling can lead to variations in diffusion. Here, a higher concentration of anions in the positive electrodes seem to be beneficial for diffusion. These results cannot be easily compared with experiments as confinement normally leads to slower diffusion~\citep{Forse17,Burt16,Asres25,Kress26}. In the future, a more accurate representation of this could be implemented through the parametrization of the systems.

\subsection{Influence of electrode structure on NMR spectra}

\begin{figure}[ht!]
    \centering    
    \includegraphics[scale=0.29]{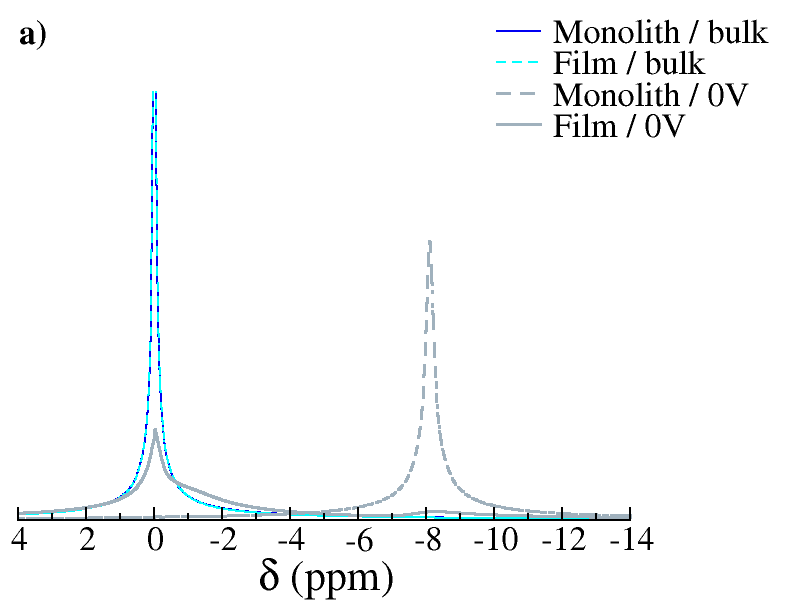}
    \includegraphics[scale=0.29]{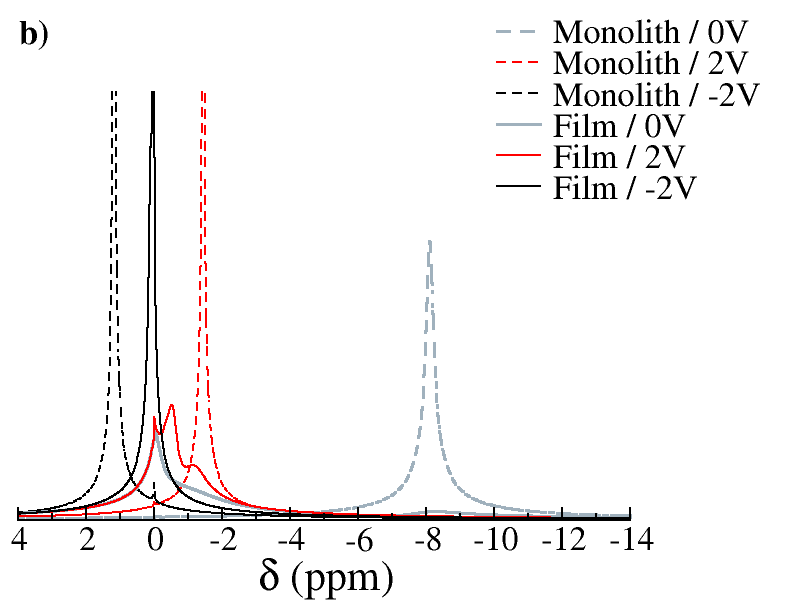}
    \caption{NMR spectra calculated for different parts of the monolith and film based supercapacitors, with various applied potentials. The residence time is the same for all lattice sites, equal to 1~ms.}
    \label{fig:NMR-mono-film}
\end{figure}

An important application of the LPC3D software is the prediction of NMR spectra to help interpret experimental results~\citep{Forse15b,anagha2021,Liu24}. We now apply it to investigate the influence of electrode morphology on NMR spectra. Figure~\ref{fig:NMR-mono-film}a shows the NMR spectra calculated for the monolith and film based supercapacitors at 0~V. As expected, the two systems show a peak at 0~ppm for the part of the system corresponding to the bulk electrolyte. For the electrode part, the monolith shows a single peak at around -8.1~ppm consistent with the most frequent pore sizes in the pore size distribution (between 0.8 and 1.2~nm). The NMR spectrum for the film electrode is more complex with a peak at around 0~ppm, a large shoulder peak at -1.2~ppm, and a small flat peak centered at -8.3~ppm. The intermediate shoulder peak at around -1.2~ppm probably corresponds to an exchange peak between bulk-like sites and particle sites in the electrode. Indeed such signatures have already been observed in other cases~\citep{lahrar2024,Lyu24}.

Figure~\ref{fig:NMR-mono-film}b shows the NMR spectra calculated for the same systems at 2~V, for the negative and positive electrodes. Again, the monolith case leads to sharp peaks shifted to higher frequencies compared to the 0~V case. This shift is consistent with \emph{in situ} NMR experiments on a large range of electrolytes~\citep{john2014,Forse21} and is explained by modifications in ring currents when the electronic density changes in the porous carbons~\citep{anagha2021}. The spectrum for the positive electrode at 2~V is quite complex with 3 peaks relatively close. This is probably due to a stronger exchange peak observed when the frequency of the anions in the pores gets closer to the frequency in the bulk (again due to ring current effects). On the contrary, the spectrum for the negative film electrode shows a single peak. In this case, the frequency for anions and particles is probably so close that the resulting peak only appears at 0~ppm. The increased diffusion in this case (seen through the diffusion coefficients calculated) also potentially affects the merging of the two bulk/particle peaks.

\subsection{Influence of ion dynamics on NMR spectra}

\begin{figure}[ht!]
    \centering    
    \includegraphics[scale=0.29]{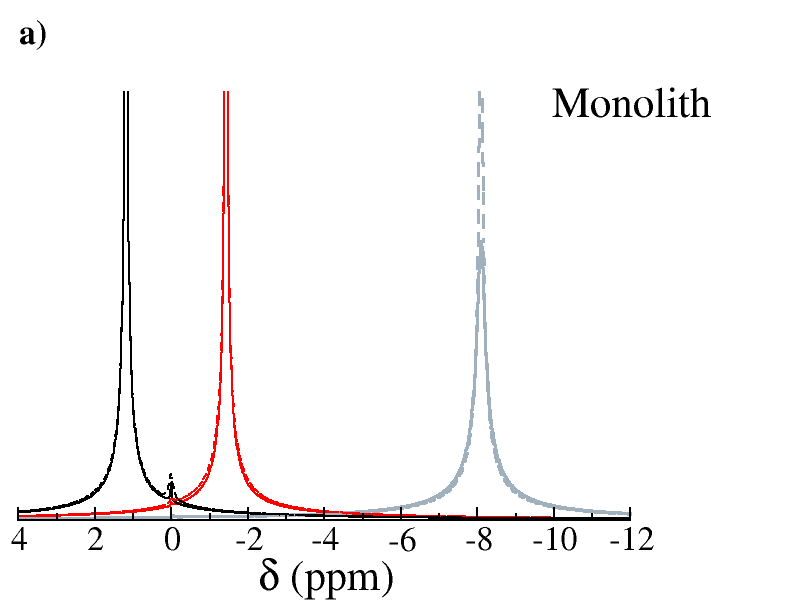}
    \includegraphics[scale=0.29]{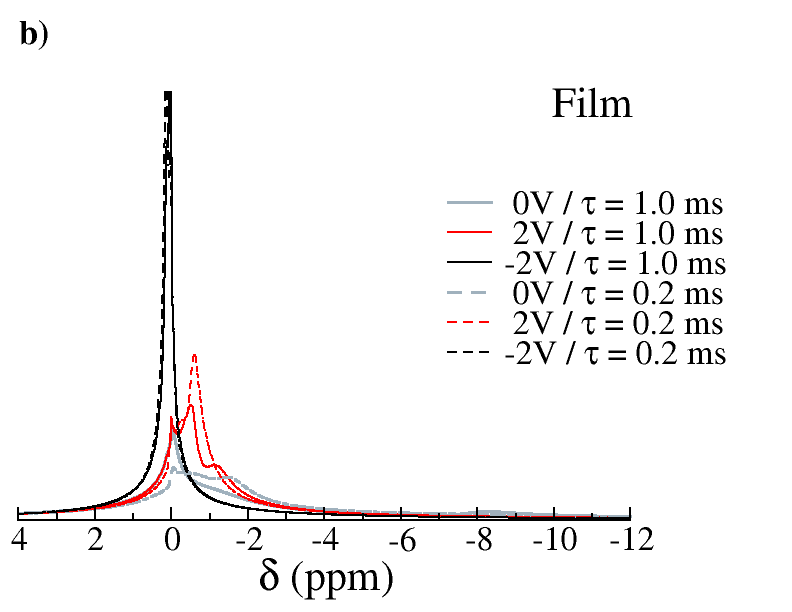}
    \caption{NMR spectra calculated for the electrodes of the monolith and film based supercapacitors, with various applied potentials. The residence time is the same for all lattice sites, equal to 1~ms or 0.2~ms as indicated.}
    \label{fig:NMR-tau}
\end{figure}

To characterize a bit more in depth the effect of ion dynamics on the resulting NMR spectra, some calculations were conducted with a residence time of 0.2~ms instead of 1~ms. For the monolith electrodes, as shown in Figure~\ref{fig:NMR-tau}a, the residence time does not affect much the spectra calculated. In this case, there is only very limited exchange between anions in the electrodes and anions in the bulk. 
For the film electrodes, as shown in Figure~\ref{fig:NMR-tau}b, the residence time influences the shape of the spectra and the position of the peaks. The effect is limited for the negative electrode with a 2~V potential but noticeable otherwise. While this change of residence time has no influence on the ion populations in the pores, the acceleration of the exchange between anions in sites having different frequencies still lead to modification in the spectrum. This suggests that absolute values of chemical shifts should be interpreted carefully and that variable temperature experiments can bring useful information regarding ion dynamics in the systems. Such variable temperature experiments have indeed already been used in the past for such purpose~\cite{Forse15,Lyu24}. It is worth noting that exploring these effects theoretically is only possible with a representation of the spatial distribution of frequencies, which is not accessible through classical analytical models.

\section{Conclusion}

In this work, we presented the new implementation and testing of a mesoscopic model, LPC3D, which is used to simulate electrolyte species diffusion in porous carbons and in full supercapacitors. The initial code, implemented serially in C, was reimplemented in C++ using {\tt{PyStencils}} framework, which generates C++ and CUDA code from the mathematical definitions. The new version of LPC3D is orders of magnitude faster and allows one to simulate systems several thousand times larger than the initial code. Performance tests show that the runtime and memory usage scale, as expected from the equations of the model, linearly with the number of sites. The strong scaling on a single HPC node is satisfactory, especially for large numbers of lattice sites.

Using this new implementation, we were able to simulate supercapacitors with two electrodes in contact with an electrolyte and investigate the influence of the electrode morphology on the properties of these systems. In particular, we have shown that monolith and film electrodes have very different NMR signatures and that simple changes in the diffusion can lead to large differences in the NMR spectra.

In the future, several routes could be followed to increase further the efficiency and/or the accuracy of the simulations. On the one hand, the input parameters are currently obtained through expensive classical molecular dynamics simulations. An alternative could be to use molecular density functional theory instead. This approach allows to sample very efficiently the phase space of polar fluids, such as water,~\cite{jeanmairet2013a} even in contact with electrode surfaces.~\cite{jeanmairet2019b} On the other hand, the parametrization of the systems could include a more precise representation of the difference in diffusion between the bulk and the electrodes / particles in which mobility is decreased due to confinement.

\section*{Author Contributions}

El Hassane Lahrar: Methodology, Software, Validation, Investigation, Formal analysis, Writing – Original Draft, Writing – Review \& Editing. Mathieu Salanne: Methodology, Writing – Review \& Editing, Resources, Funding acquisition. Rudolf Weeber: Methodology, Software, Writing – Review \& Editing. Céline Merlet: Conceptualization, Formal analysis, Methodology, Software, Validation, Writing – Original Draft, Writing – Review \& Editing, Supervision, Resources, Funding acquisition.

\section*{Conflicts of interest}

There are no conflicts to declare.

\section*{Acknowledgments}

This project has received funding from the European Union, the European High Performance Computing Joint Undertaking (JU) and countries participating in the MultiXscale project under grant agreement No 101093169. This work was granted access to the HPC resources of the CALMIP supercomputing centre under the allocation P21014, and HPC resources of the VEGA supercomputing centre under the allocation EHPC-BEN-2025B11-078. The authors acknowledge Caspar van Leeuwen for the integration of LPC3D to EESSI.

\section*{Data availability}

The data corresponding to the plots reported in this paper, as well as example input files for the MD simulations, are available in the Zenodo repository with the identifier 10.5281/zenodo.00000000.

%\bibliography{references}

\providecommand{\latin}[1]{#1}
\makeatletter
\providecommand{\doi}
  {\begingroup\let\do\@makeother\dospecials
  \catcode`\{=1 \catcode`\}=2 \doi@aux}
\providecommand{\doi@aux}[1]{\endgroup\texttt{#1}}
\makeatother
\providecommand*\mcitethebibliography{\thebibliography}
\csname @ifundefined\endcsname{endmcitethebibliography}
  {\let\endmcitethebibliography\endthebibliography}{}

\end{document}